\documentstyle[aps,prl,epsf,floats]{revtex}

\begin{document}
\draft


\twocolumn[\hsize\textwidth\columnwidth\hsize\csname@twocolumnfalse%
\endcsname
\title{Orbital Dynamics: The Origin of Anomalous Magnon
Softening in Ferromagnetic Manganites}
\author{G.\ Khaliullin$^{1,2}$ and R.\ Kilian$^1$}
\address{$^1$Max-Planck-Institut f\"ur Physik komplexer Systeme,
N\"othnitzer Strasse 38, D-01187 Dresden, Germany\\
$^2$Max-Planck-Institut f\"ur Festk\"orperforschung,
Heisenbergstrasse 1, D-70569 Stuttgart, Germany}
\date{Version date: 15 April, 1999}
\maketitle


\begin{abstract}
We study the renormalization of magnons by charge and coupled orbital-lattice 
fluctuations in colossal magnetoresistance compounds. The model considered is an
orbitally degenerate double-exchange system coupled to Jahn-Teller active 
phonons. The modulation of ferromagnetic bonds by low-energy orbital fluctuations
is identified as the main origin of the unusual softening of the zone-boundary
magnons observed experimentally in manganites. 
\end{abstract}

]


Recently, distinct new features in the spin dynamics of the 
ferromagnetic manganese oxide compound 
Pr$_{0.63}$Sr$_{0.37}$MnO$_3$ have been reported \cite{HWA98}.
Striking deviations from the predictions of the canonical 
double-exchange (DE) theory \cite{ZAG,KUB72,FUR96} were observed.
In particular, an anomalous softening of magnons at the zone
boundary was found even well below the Curie temperature $T_C$.
It is noted that these anomalies are closely related to the
reduced values of $T_C$: For higher-$T_C$ compounds no
considerable deviations from a simple cosine dispersion is
observed \cite{PER96}. 

These experimental findings seem to be of high importance.
They in fact indicate that some very specific features of
magnetism in colossal magnetoresistance manganites have still
to be identified. 
In this paper we propose a mechanism which might explain
the above experimental observations. Our basic idea is the
following: The strength of ferromagnetic interaction in 
a given bond strongly depends on which  orbital is occupied
by an $e_g$ electron (Fig.\ \ref{FIG1}).
Suppose that orbitals and Jahn-Teller (JT) distortions are disordered
in the ferromagnetic phase. It is then evident that temporal
fluctuations of orbitals affect the short-wavelength magnons
through a strong modulation of exchange bonds. Quantitatively,
the effect is expected to be controlled by the characteristic
time scale of orbital fluctuations: A slowing-down
of the dynamics of the coupled system of orbitals and JT
phonons should lead to a stronger magnon renormalization. In other
words, the observed zone-boundary magnon softening is interpreted
in this picture as a precursor effect in the proximity of static 
orbital-lattice ordering. 

More specifically, we calculate the dispersion of one-magnon
excitation at zero temperature. First, we map the ferromagnetic
Kondo-lattice model onto a Hamiltonian of interacting
magnons and spinless fermions. Yet the fermions carry orbital
quantum number, the fluctuations of which the magnon
can scatter on. Second, we show that as the orbitals are strongly
coupled to the lattice there is an indirect coupling of 
magnons to JT phonons via the orbital sector.
Put another way, the orbital-lattice coupling produces a low-energy
component in the orbital fluctuation spectrum at
phonon frequencies. We then calculate the self-energy
corrections to the magnon dispersion perturbatively  
(employing a $1/S$ expansion). Charge fluctuations
are found to produce a moderate softening of magnons throughout
the Brillouin zone. The effect of orbital and lattice 
fluctuations on the magnon dispersion is more pronounced and
becomes dramatic as static order in the orbital-lattice sector
is approached. 

We start with a model describing the ferromagnetic coupling
of double-degenerate $e_g$-band electrons to an otherwise 
noninteracting system of localized spins $\bbox{S}_t$ on a 
cubic lattice: 
\begin{eqnarray}
H &=& -\sum_{\langle ij \rangle_{\gamma}} 
t_{\gamma}^{\alpha \beta}\left(
e^{\dagger}_{i \sigma\alpha} e_{j \sigma\beta}+\text{H.c.}\right)
-J_H \bbox{S}_{it}\bbox{s}_{ie}\nonumber\\
&&+\sum_{\alpha} U n_{i\uparrow\alpha} n_{i\downarrow\alpha}
+{\sum_{\alpha\ne\beta}}' \left(U'-J_H \hat{P}\right)
n_{i\alpha} n_{i\beta}
\label{HT}
\end{eqnarray}
with $\gamma\in\{x,y,z\}$ and $\hat{P} = (\bbox{s}_{i\alpha}
\bbox{s}_{i\beta}+\frac{3}{4})$. Hereafter, the indices $\alpha/\beta$ 
and $\sigma$ stand for orbital and spin quantum numbers of $e_g$ 
electrons, respectively; summation over repeated indices is 
implied and double counting is excluded from the primed sum.
The spin operator $\bbox{s}_{i\alpha}$ acts on orbital $\alpha$ 
and $\bbox{s}_{ie}=\sum_{\alpha}\bbox{s}_{i\alpha}$.
The last two terms in Eq.\ (\ref{HT}) describe intra- (inter-)
orbital Coulomb interaction $U$ ($U'$) and Hund's coupling
between $e_g$ electrons in doubly occupied states. Correlations are 
assumed to be strong: $U, U', (U'-J_H) \gg t$. 
The important point is the peculiar orbital and bond dependencies
of the electron hopping matrix elements. In orbital basis
$\alpha \in \{|3z^2-r^2\rangle, |x^2-y^2\rangle \}$ (see Fig.\ 
\ref{FIG1}) one has \cite{KUG82}:
\begin{figure}
\noindent
\centering
\setlength{\unitlength}{0.65\linewidth}
\begin{picture}(1.0,0.583)
\put(0,0){
\epsfxsize=\unitlength
\epsffile{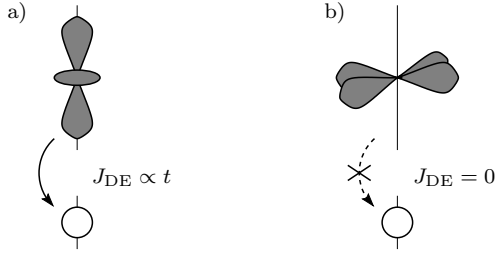}}
\put(0.15,0.16){\fontsize{0.05\unitlength}{0}\selectfont
$J_{\text{DE}}\propto t$}
\put(0.91,0.16){\fontsize{0.05\unitlength}{0}\selectfont
$J_{\text{DE}}=0$}
\put(-0.05,0.55){\fontsize{0.05\unitlength}{0}\selectfont a)}
\put(0.7,0.55){\fontsize{0.05\unitlength}{0}\selectfont b)}
\end{picture}\\[6pt]
\caption{The $e_g$-electron transfer amplitude which controls
the double-exchange interaction $J_{\text{DE}}$ strongly depends on orbital
orientation.}
\label{FIG1}
\end{figure}
\[
t_{x/y}^{\alpha\beta} = 
t\left(\begin{array}{cc}
1/4 & \mp\sqrt{3}/4\\
\mp\sqrt{3}/4 & 3/4
\end{array}\right), \quad
t_{z}^{\alpha\beta} = 
t\left(\begin{array}{cc}
1 & 0\\
0 & 0
\end{array}\right).
\]

We work in the limit $J_H\rightarrow \infty$. Then the combined
action of the on-site Hund's ``ferromagnetism'' and the 
spin-diagonal nature of electron hopping results in a global
ferromagnetic ground state. Two different contributions to the
spin stiffness should be noticed: Conventional double exchange due to 
metallic charge motion and superexchange which accounts
for the kinetic energy gain by virtual hoppings of $e_g$ electrons.
We now derive the 
effective Hamiltonian describing the spin excitations at $T\ll T_C$.

{\it Double exchange}.--- At $J_H \rightarrow \infty$, 
$\bbox{S}_t$ and the $e_g$ spin $\bbox{s}_e$ are not independent
anymore: $\bbox{s}_{ie} = n_{ie} \bbox{S}_t/2S_t$.
They form a total on-site spin $\bbox{S}_i$ with spin value
$S_t+\case{1}{2}$ if an $e_g$ electron is present,
$n_{ie}=1$, or simply $S_t$ otherwise. 
The unification of band and local spin subspaces implies 
{\it spin-charge separation}. That is, the spin component
$b_{i\sigma}$ of the $e_g$ electron $e_{i \sigma\alpha}=
b_{i \sigma}c_{i\alpha}$ is decoupled and ``absorbed'' by the total
spin, and we are left with a charged fermion $c_{i\alpha}$ 
carrying the orbital index. At $T\ll T_C$, the Schwinger boson
$b_{i\downarrow}$ simply becomes a part of the magnon operator $b_i$,
namely $b_{i\downarrow} = b_i/(2S)^{1/2}$, while
$b_{i\uparrow}$ is almost condensed, $b_{i\uparrow} = 
(n_{ie}-n_{i\downarrow})^{1/2} \approx 1-b_i^{\dagger} b_i/4S$.
Hereafter, $S=S_t+\case{1}{2}$. Further, we keep the relevant 
leading $1/S$ terms only and assume small doping $x$. The 
kinetic energy then reads
\begin{eqnarray}
H_{\text{kin}} &=& -\sum_{\langle ij \rangle_{\gamma}} 
t_{\gamma}^{\alpha \beta}c^{\dagger}_{i\alpha} c_{j\beta}
+\frac{1}{2S}\sum_{\langle ij \rangle_{\gamma}}t_{\gamma}^{\alpha \beta}
\nonumber\\&&\times
c^{\dagger}_{i\alpha} c_{j\beta} \Big(\frac{1}{2}
b^{\dagger}_i b_i + \frac{1}{2} b^{\dagger}_j b_j
-b^{\dagger}_i b_j \Big) + \text{H.c.}
\label{HKI}
\end{eqnarray} 
The first term here describes the fermionic motion in a ferromagnetic
background (yet this motion is strongly correlated in the presence of 
orbital disorder \cite{ISH97,KIL98}),
while the second term controls the spin dynamics and spin-fermion
interaction. The latter term would come out from a spin-wave expansion of
an effective Heisenberg model
$J_{\text{DE}}(\bbox{S}_i \bbox{S}_j)$ with
$J_{\text{DE}}= (2S^2)^{-1} t_{\gamma}^{\alpha\beta}\langle c^{\dagger}_{i\alpha}
c_{j\beta} \rangle$ if one considers the fermionic 
sector on {\it average} as in mean-field treatments of the DE model. However, 
the bond variable $c^{\dagger}_{i\alpha} c_{j\beta}$ is a fluctuating complex 
quantity and the spin structure of Eq.\ (\ref{HKI}) is therefore not of
Heisenberg form. Explicitly ($T\ll T_C$):
\begin{eqnarray}
H_{\text{kin}} &=& -\sum_{\langle ij \rangle_{\gamma}} 
t_{\gamma}^{\alpha \beta} c^{\dagger}_{i\alpha} c_{j\beta}\nonumber\\
&&\times\Big[\frac{3}{4} + \frac{1}{4S^2} \left(S_i^z S_j^z
+ S_i^- S_j^+\right)\Big] + \text{H.c.}
\label{HKI2}
\end{eqnarray}
In the classical limit for spins, one recovers from Eq.\ (\ref{HKI2}) 
the effective fermionic model with phase-dependent hopping 
\cite{MUL96,MIL95,NOTE1}.
In addition to the above peculiarities, the band/local duality
of spin in the DE system further results in the following important
point: Going to momentum representation in Eq.\ (\ref{HKI}) one
immediately realizes that the magnon-fermion interaction vertex does
not vanish at zero-momentum transfer. What is wrong? The boson 
$b_i$ should not be regarded as a physical magnon. The true Goldstone
particle of the DE model is the following object with both local
spin and itinerant features:
\begin{eqnarray}
B_i &=& b_i\Big[n_{ic} + \sqrt{\frac{2S-1}{2S}}
\left(1-n_{ic}\right)\Big]\nonumber\\
&\approx & b_i - \frac{1}{4S} \left(1-n_{ic}\right)b_i.
\label{BI}
\end{eqnarray}
The composite character of the physical magnon $B_i$ is the price
one has to pay for spin-charge separation and for the rearrangement
of the original Hilbert space. The itinerant component of $B_i$
is of order $1/S$ only. However, the spin stiffness is itself of the
same order, and the $1/S$ correction in Eq.\ (\ref{BI}) 
is of crucial importance to ensure spin dynamics 
consistent with the Goldstone theorem. Now, commuting Eq.\ (\ref{BI})
with Eq.\ (\ref{HKI}) one finds the mean-field magnon dispersion
$\omega_{\bbox{p}}$ and the correct momentum structure of the magnon-fermion
scattering vertex: 
\begin{eqnarray}
\left[B_{\bbox{p}},H\right] &=&
\omega_{\bbox{p}} B_{\bbox{p}}+ \frac{t}{2S} \sum_{\bbox{q}}
A_{\bbox{p}}^{\alpha\beta}(\bbox{k})
c^{\dagger}_{\bbox{k}\alpha}c_{\bbox{k}-\bbox{q},\beta}B_{\bbox{p}+\bbox{q}},
\label{COM}\\
&&A_{\bbox{p}}^{\alpha\beta}(\bbox{k}) = \gamma_{\bbox{k}}^{\alpha\beta}
-\gamma_{\bbox{k}+\bbox{p}}^{\alpha\beta}.\nonumber
\end{eqnarray}
Here, $\omega_{\bbox{p}} = zD_s\left(1-\gamma_{\bbox{p}}\right)$,
$D_s = J_{\text{DE}} S$, $J_{\text{DE}}$ is defined above, $z=6$, 
and the form factors $\gamma_{\bbox{k}} = z^{-1}\sum_{\bbox{\delta}}
\exp(i\bbox{k}\bbox{\delta})$, $\gamma_{\bbox{k}}^{\alpha\beta}
= (zt)^{-1} \sum_{\bbox{\delta}} t^{\alpha\beta}_{\delta} 
\exp(i\bbox{k}\bbox{\delta})$. The physics behind the second term
in Eq.\ (\ref{COM}) is the temporal fluctuations of the ``exchange
constant'' due to the charge and orbital dynamics. From
now on, the conventional diagrammatic method with bare magnon-fermion
vertex given in Eq.\ (\ref{COM}) can be used.

Now we turn to the correlations in the fermionic band. We separate
the {\it charge} and {\it orbital} degrees of freedom by the
parameterization $c_{i\alpha} = h^{\dagger}_i f_{i\alpha}$, where
the slave boson $h_i$ and the fermion $f_{i\alpha}$ represent the 
density and orbital fluctuations, respectively \cite{KIL98}. 
Further, approximating $\delta(c^{\dagger}_{i\alpha}c_{j\beta})
= \langle h^{\dagger}_j h_i\rangle f^{\dagger}_{i\alpha}
f_{j\beta} + \langle f^{\dagger}_{i\alpha} f_{j\beta}\rangle 
h^{\dagger}_j h_i$, one finds from Eq.\ (\ref{COM}) two different 
contributions to the magnon self energy.

{\it Superexchange}.---
At low dopings, the virtual charge transfer across the Hubbard
gap becomes of importance. In limit of large $J_H$,
transitions to the high-spin intermediate state with energy 
$U_1=(U'-J_H) \ll U',U$ dominate. The 
corresponding superexchange Hamiltonian is then obtained to be:
\begin{equation}
H_{\text{SE}} = -J_{\text{SE}} \sum_{\langle ij\rangle_{\gamma}} 
\left[\bbox{S}_i \bbox{S}_j + S(S+1)\right]
\left(1-\tau_i^{\gamma} \tau_j^{\gamma}\right).
\label{HSE}
\end{equation}
Here, $J_{\text{SE}} = (t^2/U_1)[2S(2S+1)]^{-1}$ and
$\tau^{x/y} = \left(\sigma^z\pm\sqrt{3}\sigma^x\right)/2$, $\tau^{z} = 
\sigma^z$ with  Pauli matrices $\sigma^{x/z}$ acting on orbital subspace
$f_{\alpha}$. Eq.\ (\ref{HSE}) is the generalization of orbitally degenerate
superexchange models \cite{KUG82,OLE99} for arbitrary values of spin.
Superexchange is of {\it ferromagnetic} nature 
because the $J_H\rightarrow \infty$ limit is assumed, and the
exchange strength depends on orbital orientations \cite{KHA97}. In an orbitally
disordered state one can represent the orbital part of Eq.\ (\ref{HSE})
in terms of fluctuating bond operators $f^{\dagger}_{i\alpha}
f_{j\beta}$ \cite{KIL98}. Then the effect of superexchange is
simply to add a new term
\begin{equation}
A_{\bbox{p}}^{\alpha\beta}(\bbox{k},\bbox{q}) = x_0\left(
\gamma_{\bbox{k}}^{\alpha\beta}
+\gamma_{\bbox{k}-\bbox{q}}^{\alpha\beta}
-\gamma_{\bbox{k}+\bbox{p}}^{\alpha\beta}
-\gamma_{\bbox{k}-\bbox{q}-\bbox{p}}^{\alpha\beta}\right)
\end{equation}
to the scattering amplitude in Eq.\ (\ref{COM}) and to renormalize 
$D_s \rightarrow (J_{\text{DE}} + J_{\text{SE}}) S =
t\chi_0(x+x_0)/2S$. Here the constant 
$x_0 = 2\chi_0 t/U_1$ with mean-field parameter 
$\chi_0 = \langle f^{\dagger}_{i+c}f_i\rangle\approx\case{1}{2}(1-x)$
defines the effective doping level below which the superexchange
contribution becomes of importance. 

To summarize up to now: We have developed a quantum theory of the
double-exchange model at $T< T_C$ which also
includes the correlation effects present in transition metal oxides.
The obvious advantage of this approach over earlier treatments 
\cite{KUB72,FUR96} is that the theory has a transparent structure
adjusted to describe the low-energy spin dynamics, and the large atomic
scale $J_H$ does not appear in the calculations, either. 
Yet the corrections $t/J_H$ can be accounted for perturbatively,
if necessary. 

\begin{figure}
\noindent
\centering
\epsfxsize=0.8\linewidth
\epsffile{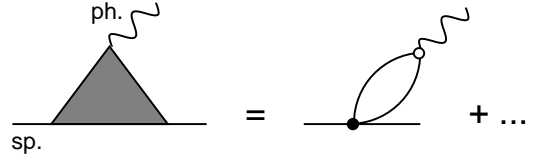}\\[6pt]
\caption{Effective spin-phonon coupling. Perturbatively,
it is controlled by the orbital susceptibility (fermion
bubble), Jahn-Teller (open dot $\propto g_0$) and
double-exchange (filled dot $\propto t$) interaction vertices.}
\label{FIG2}
\end{figure}

{\it Indirect magnon-phonon coupling via orbital sector}.--- 
An important piece of physics relevant to the experiment \cite{HWA98}
is still missed in model (\ref{HT}), that is the Jahn-Teller
orbital-lattice coupling \cite{MIL96}. This can be written as:
\begin{equation}
H_{\text{JT}} = -g_0\left(Q_{2i}\sigma_i^x+Q_{3i}\sigma_i^z\right),
\end{equation}
where $Q_2$ and $Q_3$ are the distortions of appropriate symmetry
\cite{KUG82}. The deformation energy including intersite correlations
is given by
\begin{equation}
H_{\text{ph}} = \frac{1}{2} K\sum_i \bbox{Q}_i^2 + K_1\sum_{\langle ij
\rangle_{\gamma}}\tilde{Q}_{3i}^{\gamma} \tilde{Q}_{3j}^{\gamma},
\label{DEF}
\end{equation}
with $\tilde{Q}_3^{x/y} = (Q_3\pm \sqrt{3}Q_2)/2$, $\tilde{Q}_3^z = 
Q_3$, and $\bbox{Q} = (Q_2,Q_3)$. In general, the JT interaction
strongly mixes the orbital and lattice dynamics, leading to a fluctuation
of exchange bonds at low (phonon) frequencies. Here, we treat this
problem only perturbatively as shown in Fig.\ \ref{FIG2}. 
We arrive at the following effective spin-phonon Hamiltonian:
\begin{equation}
H_{\text{s-ph}} = -\sum_{\bbox{p}\bbox{q}} \left(
\bbox{g}_{\bbox{p}\bbox{q}} \bbox{Q}_{-\bbox{q}}\right)
B^{\dagger}_{\bbox{p}} B_{\bbox{p}+\bbox{q}}.
\end{equation}
The coupling constants $g^{(\alpha)}_{\bbox{p}\bbox{q}} = 
g_0 a_0(\eta^{(\alpha)}_{\bbox{q}} - \eta^{(\alpha)}_{\bbox{p}}-
\eta^{(\alpha)}_{\bbox{p}+\bbox{q}})/S$, where 
$\eta^{(2)}_{\bbox{q}} = -\sqrt{3}(c_x-c_y)/2$,
$\eta^{(3)}_{\bbox{q}} = c_z-\case{1}{2}c_x-\case{1}{2}c_y$,
$c_{\alpha} \equiv \cos q_{\alpha}$, and the parameter
$a_0 = t(x+x_0)\langle (f^{\dagger}_{i+c} f_i)(\sigma_i^z)\rangle_{\omega=0}$.

\begin{figure}
\noindent
\begin{minipage}{0.3\linewidth}
\epsfxsize=\linewidth
\epsffile{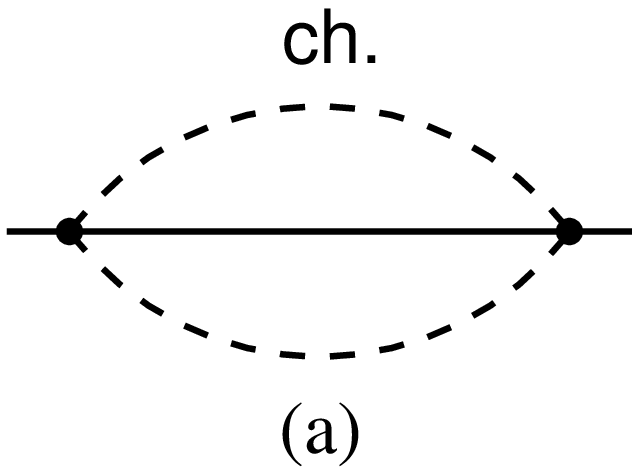}
\end{minipage}
\hfill
\begin{minipage}{0.3\linewidth}
\epsfxsize=\linewidth
\epsffile{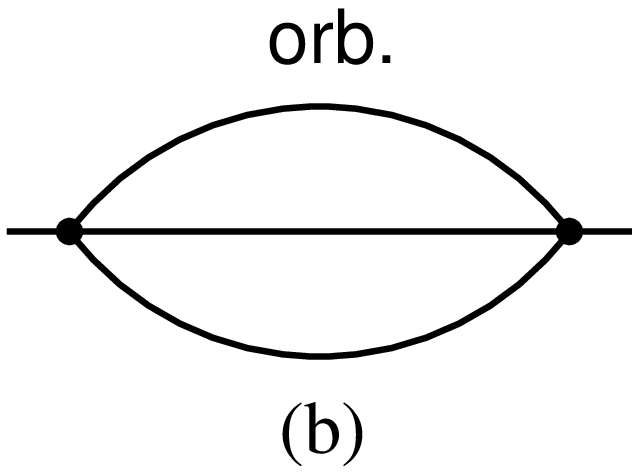}
\end{minipage}
\hfill
\begin{minipage}{0.3\linewidth}
\epsfxsize=\linewidth
\epsffile{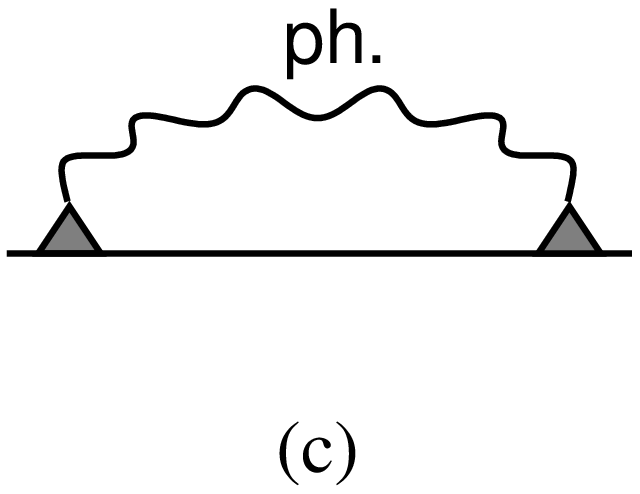}
\end{minipage}\\[6pt]
\caption{Magnon self energies.}
\label{FIG3}
\end{figure}

\begin{figure}
\noindent
\centering
\epsfxsize=0.8\linewidth
\epsffile{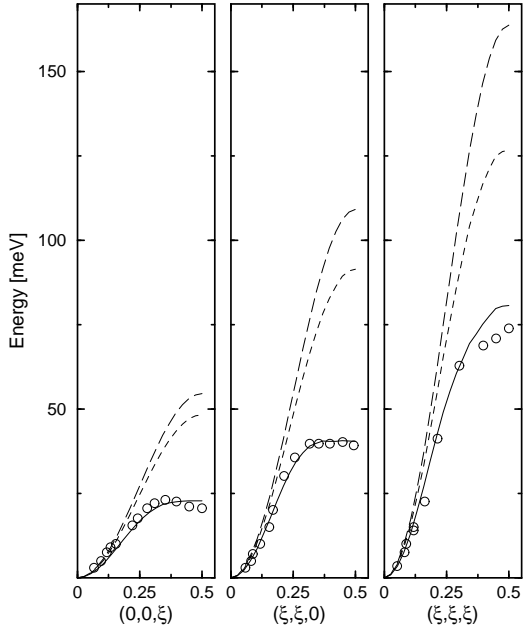}
\caption{Magnon dispersion along $(0,0,\xi)$, 
$(\xi,\xi,0)$, and $(\xi,\xi,\xi)$ directions, where 
$\xi=0.5$ at the zone boundary. Solid lines represent
the theoretical result including charge, orbital, and
lattice effects and are fitted to experimental data 
\protect\onlinecite{HWA98} denoted by circles.
For comparison the bare dispersion and the one including 
only charge effects are indicated by long-dashed and dashed lines,
respectively. The parameter $k_1=-0.33$.}
\label{FIG4}
\end{figure}

\begin{figure}
\noindent
\centering
\epsfxsize=0.8\linewidth
\epsffile{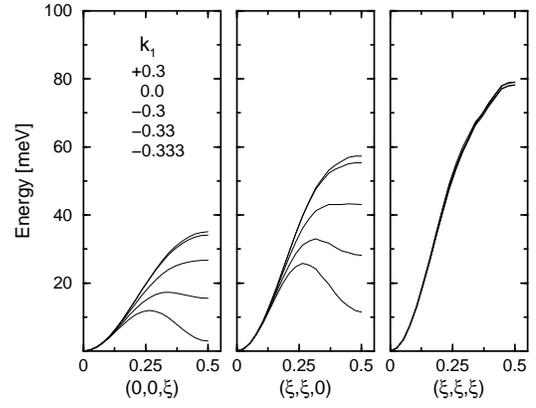}
\caption{Magnon dispersion for different values of $k_1$.
$E_{\text{JT}}a_0^2 = 0.006$~eV is used. The softening enhances
as $k_1\rightarrow -\case{1}{3}$ corresponding to 
ferro-type orbital-lattice order.}
\label{FIG5}
\end{figure}

{\it Magnon self energies}--- 
We are now in the position to calculate the renormalization of the magnon 
dispersion. The leading $1/S^2$ corrections are shown in Fig.\
\ref{FIG3}.
Charge and orbital susceptibilities in Figs.\ \ref{FIG3}(a) and
\ref{FIG3}(b), respectively, are calculated using mean-field Green's functions
in slave boson $h$ and fermion $f$ subspaces. For the spectral density of JT 
phonons in Fig.\ \ref{FIG3}(c) we use
\begin{equation}
\rho^{\text{ph}}_{\pm}(\omega,\bbox{q}) = 
\frac{1}{\pi}\frac{\omega}{\omega_{\bbox{q}}^{\pm}}
\frac{\Gamma}{(\omega-\omega_{\bbox{q}}^{\pm})^2+\Gamma^2}
\end{equation}
accounting phenomenologically for the damping $\Gamma$ due to coupling 
to orbital fluctuations. The phonon dispersions
$\omega_{\bbox{q}}^{\pm} = \omega_0^{\text{ph}}[\kappa_{1\bbox{q}}
\pm(\kappa_{2\bbox{q}}^2+\kappa_{3\bbox{q}}^2)^{1/2}]^{1/2}$ with
$\omega_0^{\text{ph}} = (K/M)^{1/2}$
follow from Eq.\ (\ref{DEF}). Here, $\kappa_{1\bbox{q}} = 1+
k_1(c_x+c_y+c_z)$, $\kappa_{2\bbox{q}} = k_1 \eta^{(2)}_{\bbox{q}}$,
$\kappa_{3\bbox{q}} = k_1\eta^{(3)}_{\bbox{q}}$, and $k_1 = K_1/K$.
The expressions obtained from the diagrams in Fig.\ \ref{FIG3} contain 
summations over momentum space which we perform numerically. We find
the effect of charge fluctuations on the magnon spectrum to be quite 
featureless and moderate (see Fig.\ \ref{FIG4}),
which is due to the fact that the spectral density of charge fluctuations
lies well above the magnon band. On
contrary, the relatively low-energy fluctuations of the orbital
and lattice degrees of freedom are found to affect the
spin-wave dispersion in a peculiar way, particularly in $(0,0,q)$ and
$(0,q,q)$ directions. Fig.\ \ref{FIG4} shows the magnon
dispersion. The hopping amplitude $t=0.4$~eV is chosen to fit the
spin stiffness in Pr$_{0.63}$Sr$_{0.37}$MnO$_3$; further we use 
$U_1=4$~eV \cite{OLE99}. The phonon contribution 
depends on the quantity $(g_0a_0)^2/2K\equiv E_{\text{JT}}a_0^2$. We set
$E_{\text{JT}}a_0^2=0.004$~eV \cite{REM1}, $\omega_0^{\text{ph}} = 0.08$~eV 
\cite{REM2}, $\Gamma=0.04$~eV.
Our key observation is the crucial effect of intersite
correlations of JT distortions, controlled by $k_1$, on the
magnon dispersion (see Fig.\ \ref{FIG5}).
To explain the experimental data \cite{HWA98} we are forced to assume
{\it ferro-type} correlations ($k_1<0$). We interpret this surprising 
result in the following way: Conventionally $k_1>0$ corresponding
to AF order of JT distortions and orbitals is expected in
undoped compounds \cite{KUG82}.
However, in the doped case charge mobility prefers a
ferro-type local orientation of orbitals which minimizes the 
kinetic energy. 
This competition between JT and kinetic energies can be simulated
by tuning $k_1$. At large enough doping, ferro-type orbital 
correlations are expected to prevail, hence effectively $k_1<0$.
In fact, a ferro-type ordering of orbitals
leading to a layered AFM spin structure is experimentally observed
\cite{KAW97,MOR98} at dopings about $x=0.5$. As this instability is 
approached, low-energy, quasi static fluctuations of exchange bonds
develop, yielding a magnon evolution as shown in Fig.\ \ref{FIG5}.
Remarkably, the small $q$ spin stiffness is not affected
by this physics, while a strong reduction of $T_C$ by soft magnons at the zone
boundary is predicted. This explains the origin of the anomalous enhancement
of the $D_s/T_C$ ratio in low-$T_C$ manganites \cite{FER98}.

In summary, we have presented a theory of spin dynamics in a model
relevant to manganites, emphasizing particularly the interplay
between double-exchange physics and orbital-lattice dynamics. 
The unusual magnon dispersion in lower-$T_C$ compounds is explained
as due to the proximity to orbital-lattice ordering. Apparently,
strongly correlated orbital fluctuations play a crucial role in the
physics of manganites.


We would like to thank H.\ Y.\ Hwang for stimulating
discussions.



\begin{references}

\bibitem{HWA98} H.\ Y.\ Hwang {\it et al.}, Phys.\ Rev.\ Lett.\
{\bf 80}, 1316 (1998).

\bibitem{ZAG} 
C.\ Zener, Phys.\ Rev.\ {\bf 82}, 403 (1951); 
P.\ W.\ Anderson and H.\ Hasegawa, {\it ibid.} {\bf 100}, 675 (1955);
P.-G.\ de Gennes, {\it ibid.} {\bf 118}, 141 (1960).

\bibitem{KUB72} K.\ Kubo and N.\ Ohata, J.\ Phys.\ Soc.\ Jpn.\
{\bf 33}, 21 (1972).

\bibitem{FUR96} N.\ Furukawa, J.\ Phys.\ Soc.\ Jpn.\
{\bf 65}, 1174 (1996).

\bibitem{PER96} T.\ G.\ Perring {\it et al.}, Phys.\ Rev.\ Lett.\ 
{\bf 77}, 711 (1996).

\bibitem{KUG82} K.\ I.\ Kugel and D.\ I.\ Khomskii,
Sov.\ Phys.-Usp.\ {\bf 25}, 231 (1982).

\bibitem{ISH97} S.\ Ishihara, M.\ Yamanaka, and N.\ Nagaosa, Phys.\ Rev.\ B
{\bf 56}, 686 (1997).

\bibitem{KIL98} R.\ Kilian and G.\ Khaliullin, Phys.\ Rev.\ B
{\bf 58}, R11~841 (1998).

\bibitem{MUL96} E.\ M\"uller-Hartmann and E.\ Dagotto,
Phys.\ Rev.\ B {\bf 54}, R6819 (1996).

\bibitem{MIL95} A.\ J.\ Millis, P.\ B.\ Littlewood, and 
B.\ I.\ Shraiman, Phys.\ Rev.\ Lett.\ {\bf 74}, 5144 (1995).

\bibitem{NOTE1} It is noticed that the phase-dependent part of the
fermionic hopping is only of order $1/S$ (see Eq.\ (\protect\ref{HKI})).

\bibitem{OLE99} L.\ F.\ Feiner and A.\ M.\ Ole\'s, Phys.\ Rev.\ B 
{\bf 59}, 3295 (1999).

\bibitem{KHA97} The effect of superexchange-bond fluctuations on the magnon
spectrum in the antiferromagnetic Kugel-Khomskii model was recently 
studied in G.\ Khaliullin and V.\ Oudovenko, Phys.\ Rev. B {\bf 56}, 
R14~243 (1997).

\bibitem{MIL96} A.\ J.\ Millis, b.\ I.\ Shraiman, and R.\ Mueller,
Phys.\ Rev.\ Lett.\ {\bf 77}, 175 (1996).

\bibitem{REM1} A mean-field calculation gives $a_0\approx 0.1$. Our
fitting then implies a reasonable Jahn-Teller binding energy
$E_{\text{JT}} \approx 0.4$~eV.

\bibitem{REM2} This number is consistent with optical reflectivity
data in Y.\ Okimoto {\it et al.}, Phys.\ Rev.\ Lett.\ {\bf 75},
109 (1995).
 
\bibitem{KAW97} H.\ Kawano {\it et al.}, Phys.\ Rev.\ Lett.\
{\bf 78}, 4253 (1997).

\bibitem{MOR98} Y.\ Moritomo {\it et al.}, Phys.\ Rev.\ B
{\bf 58}, 5544 (1998).

\bibitem{FER98} J.\ A.\ Fernandez-Baca {\it et al.},
Phys.\ Rev.\ Lett.\ {\bf 80}, 4012 (1998).

\end{references}
\end{document}